\documentclass[aps,prl,showpacs,amsmath,amsfonts,amssymb,superscriptaddress]{revtex4} 
\usepackage{psfig}
\usepackage[dvips]{graphicx}
\usepackage{isolatin1}

\newcommand{\beq}{\begin{equation}} 
\newcommand{\eeq}{\end{equation}}
\newcommand{\bea}{\begin{eqnarray}} 
\newcommand{\eea}{\end{eqnarray}} 
 
\input epsf 
 
\begin{document} 
 
\title{Curvature of the energy landscape and folding of model proteins} 
 
\author{Lorenzo N.\ Mazzoni} 
\email{mazzoni@fi.infn.it} 
\affiliation{Dipartimento di Fisica and Centro per lo Studio
delle Dinamiche Complesse (CSDC), Universit\`a di 
Firenze, via G.~Sansone 1, I-50019 Sesto Fiorentino (FI), Italy}  
 
\author{Lapo Casetti} 
\email{lapo.casetti@unifi.it} 
\thanks{Author to whom correspondence should be sent.}
\affiliation{Dipartimento di Fisica and Centro per lo Studio
delle Dinamiche Complesse (CSDC), Universit\`a di 
Firenze, via G.~Sansone 1, I-50019 Sesto Fiorentino (FI), Italy}  
\affiliation{Istituto Nazionale di Fisica Nucleare (INFN), Sezione di 
Firenze, Italy}  

\date{October 24, 2006} 
 
\begin{abstract} 
We study the geometric properties of the energy landscape of coarse-grained,
off-lattice models of polymers by endowing the configuration space with a
suitable metric, depending on the potential energy function, such that the
dynamical trajectories are the geodesics of the metric. Using numerical
simulations, we show that the fluctuations of the curvature clearly mark the
folding transition, and that this quantity allows to distinguish between
polymers having a protein-like behavior (i.e., that fold to a unique
configuration) and polymers which undergo a hydrophobic collapse but do not
have a folding transition. These geometrical properties are defined by the
potential energy without requiring any prior knowledge of the native
configuration.
\end{abstract} 
 
  
\maketitle 

Protein folding is one of the most fundamental and challenging open questions
in molecular biology. Proteins are polypeptides, i.e., polymers made of
aminoacids, and since the
pioneering experiments by Anfinsen and coworkers \cite{Anfinsen} it has been
known that the sequence of aminoacids uniquely determines the native state, 
i.e., the compact configuration the protein assumes in physiological conditions
and which makes it able to perform its biological tasks \cite{proteinbook}. To
understand how the information contained in the sequence is translated into the
three-dimensional native structure is the core of the protein folding problem,
and its solution would allow one to predict a protein's structure from the sole
knowledge of the aminoacid sequence: moreover, solving the protein folding
problem would make it possible to engineer proteins which fold to any given
structure (what is commonly referred to as the inverse folding problem), which
in turn would mean a giant leap in drug design. Despite many remarkable
advances in the last decades \cite{proteinbook}, the protein folding problem is
still far from a solution.

Within the folding problem, a basic issue stems from the observation that 
not all polypeptides are proteins: only a very small subset of all the possible
sequences of the twenty naturally occurring aminoacids have been selected by
evolution. According to our present knowledge, all the naturally selected
proteins fold to a uniquely determined native state, but a generic polypeptide
does not. Then, what makes a protein different from a generic polypeptide?
or, in other words, which are the properties a polypeptide must have to behave
like a protein, i.e., to fold into a unique native state regardless of the
initial conditions, when the environment is the correct one? Answering this
question would not directly yield a solution of the folding problem, nonetheless
it would indicate which are the minimal common properties of those polymers
which fold like a protein. To this end, 
the {\em energy landscape} picture has emerged as crucial.
Energy landscape, or more precisely potential energy landscape, is the name
commonly given to the potential energy of interaction between the
microscopic degrees of freedom of the system \cite{Wales}. 
Before having been applied to biomolecules, this concept has
proven useful in the study of other complex systems, especially of supercooled
liquids and of the glass transition \cite{naturecomplex}. The basic idea is
very simple, yet powerful: if a system has a rugged, complex energy landscape,
with many minima and valleys separated by barriers of different height, its
dynamics will experience a variety of time scales, with oscillations in the
valleys and jumps from one valley to another\footnote{This is the picture of a
classical dynamics at a finite temperature, where in addition to $\nabla V$ 
there is a stochastic
force proportional to the temperature, which makes the thermally activated
jumps between valleys possible. However, for a sufficiently large system also a
completely deterministic dynamics over the same landscape would give a similar
overall behavior.}. Then one can try to link special features of the behavior of
the system (i.e., the presence of a glass transition or the separation of time
scales) to special properties of the landscape, like the topography
of the basins around minima or the energy distribution of minima and saddles
connecting them. Anyway, a complex landscape yields a complex
dynamics, where the system is very likely to remain trapped in different
valleys when the temperature is not so high. This is consistent with a glassy
behavior, but a protein does not show a glassy behavior, it rather has
relatively low frustration. This means that there must be some property of the
landscape such to avoid too much frustration. This property is commonly
referred to as the {\em folding funnel} \cite{funnel}: though locally rugged,
the low-energy part of the energy landscape is supposed to have an overall
funnel shape so that most initial conditions are driven towards the correct
native state. Moreover, the dynamics must do that efficiently, or the protein
would not fold in reasonable times; in other words, it must be ``sufficiently
unstable'' to make trapping in local minima very unlikely, and 
saddles must efficiently connect non-native minima with the native state. 
However, a direct visualization of the energy landscape is
impossible due to its high dimensionality, and its detailed properties must be
inferred indirectly. A possible strategy is a local one: one searches for the
minima of the landscape and then for the saddles connecting different minima.
Although straightforward in principle, 
this is practically unfeasible for accurate all-atom potential energies, but
may become accessible for minimalistic potentials\footnote{Searching for {\em
all} the minima and {\em all} the saddles is impossibile even for very simple
potentials, unless it can be done analytically, because no algorithm is
available which is able to find {\em all} the solutions of $N$ coupled
nonlinear equations \protect\cite{numrec}; nonetheless, one expects that with a
considerable numerical effort a reasonable sampling of these points can be
achieved for not-too-detailed potentials, as it happens for binary
Lennard-Jones fluids \protect\cite{Grigera}.}. Minimalistic models are those
where the polymer is described at a coarse-grained level, as a chain  of $N$
beads where $N$ is the number of aminoacids; no explicit water molecules are
considered and the solvent is taken into account only by means of effective
interactions among the monomers. Minimalistic models can be relatively simple,
yet in some cases yield very accurate results which compare well with
experiments \cite{Clementi}. The local properties of the energy landscape of
minimalistic models have been recently studied 
(see e.g.\ Refs.\ \cite{Miller_Wales,Bongini1,Bongini2}) 
and very interesting clues about
the structure of the folding funnel and the differences between protein-like
heteropolymers and other polymers have been found: in particular, it has 
been shown that a funnel-like structure is present also in homopolymers, 
but what makes a big
difference is that in protein-like systems jumps between minima corresponding
to distant configurations are much more favoured dynamically \cite{Bongini2}.

The above mentioned local strategy to analyze energy landscapes requires
however a huge computational effort if one wants to obtain a good sampling. So
the following question naturally arises: is there some {\em global} property of
the energy landscape which can be easily computed numerically as an average
along dynamical trajectories and which is able to identify polymers having a
protein-like behavior? The main issue of the present Letter is to show that
such a quantity indeed exists, at least for the minimalistic model we
considered, and that it is of a geometric nature. In particular, we will show
that the fluctuations of a suitably defined curvature of the energy landscape
clearly mark the folding transition while do not show any remarkable feature
when the polymer undergoes a hydrophobic collapse without a preferred native
state. This is at variance with thermodynamic global observables, like the
specific heat, which show a very similar behavior in the case of a folding
transition and of a simple hydrophobic collapse. 

It is a classic result of analytical dynamics that the stability properties of
the trajectories of a dynamical system are completely determined by the {\em
curvature} of a suitable manifold, i.e., of the configuration space endowed
with a metric tensor $g$ depending on the potential energy $V(q_1,\ldots,q_N)$
such that its geodesics\footnote{Geodesics are the ``straight'' lines on 
a curved manifold, i.e., the curves whose velocity
vector has a vanishing covariant derivative \protect\cite{geometry}.} 
coincide with the dynamical trajectories \cite{Arnold}. Locally, a positive
curvature implies stability, while negative curvatures are associated to
instability: accordingly, the metric $g$ is such that close to minima of $V$
the curvature is positive, while saddles have negative curvatures, at
least along some direction. However, instability can be generated also by the
bumpiness of the manifold: if the curvature fluctuates along a geodesic 
it may destabilize it even without assuming 
negative values, the degree of instability being related to the size of
the fluctuations \cite{prl95pre96} (see Ref.\ \cite{physrep} for a
review). 

The geometrization of the dynamics is not unique: a particularly
convenient procedure was introduced by Eisenhart \cite{Eisenhart} by
considering an enlarged $(N+2)$-dimensional configuration space. In
terms of the coordinates $q^0,q^1,\ldots,q^N,q^{N+1}$, where
$q^1,\ldots,q^N$ are the lagrangian coordinates and $q^0$ and $q^{N+1}$ two
extra coordinates, the nonzero components of the Eisenhart metric tensor 
are (we set the masses of the particles equal to 1 for simplicity)  $g_{00} = -2
V(q)$ and $g_{ii} = g_{0\, N+1} = g_{N+1\, 0} = 1$ ($i = 1,\ldots,N$);
one can prove that the geodesics of the Eisenhart metric project
onto dynamical trajectories. The mathematical object which contains all the
information on the curvature is the curvature tensor $R$ \cite{geometry}: in the
case of the Eisenhart metric it turns out to be very simple, for its nonzero
components are given by the Hessian of the potential $V$,
$R_{0i0j} = \partial_i \partial_j V$. The curvature of the Eisenhart metric is
then just the curvature of the energy landscape itself, as a function of
$(q^1,\ldots,q^N)$. The quantity which
actually determines the stability properties of a
geodesic of velocity $v$ (in a given direction $w\perp v$) 
is the sectional curvature  
$K(v,w) = R_{ijkl}v^i w^j v^k w^l/|v \wedge w|^2$ \cite{geometry}. 
In $N$ dimensions there are $N-1$ independent $w$ directions, nonetheless the
most important information is already contained in the average of $K(v,w)$ over
the $N-1$ possible directions of $w$ \cite{physrep}. This scalar 
quantity is called the Ricci
curvature and is given by $K_R(v) = R_{ij}v^i v^j$, where $R_{ij} = R^k_{~ikj} =
g^{kl}R_{likj}$ are the components of the Ricci tensor \cite{geometry}. In the
case of the Eisenhart metric, the Ricci curvature along the direction of 
the velocity vector (i.e., the Ricci curvature ``felt'' by the system during
its motion, and which we will refer to simply as $K_R$ dropping the dependence
on $v$) is nothing but the Laplacian of the potential  \cite{physrep}, 
\beq
K_R = \triangle V~,
\label{ricci}
\eeq
i.e., the average curvature of the energy landscape. 
We may then expect that the statistical distribution of $K_R$
on the configuration space contains relevant information on the 
stability of a generic
trajectory: such an observable is then a good
candidate for a global quantity able to catch some of the features of the
landscape which characterize a protein-like behavior. 

We sampled the value of the Ricci curvature $K_R$ along the dynamical
trajectories of a minimalistic model originally introduced by Thirumalai and
coworkers \cite{Thirumalai}, a three-dimensional off-lattice model of a
polypeptide which has only three different
kinds of aminoacids: polar (P), hydrophobic (H) and neutral (N). The potential
energy is 
\beq 
V = V_\text{B} +
V_\text{A} +
V_\text{D} + V_\text{NB}
\eeq 
where 
\bea
V_\text{B} & = & \sum_{i=1}^{N-1} \frac{k_r}{2} (|{\vec{r}}_i - {\vec{r}}_{i-1}|
- a)^2\,;\\
V_\text{A} & = & \sum_{i=1}^{N-2} \frac{k_\vartheta}{2} (|{\vartheta}_i -
{\vartheta}_{i-1}| - \vartheta_0)^2\,; \\
V_\text{D} & = & \sum_{i=1}^{N-3} \left\{A_i [1 + \cos \psi_i] + B_i[1 +
\cos(3\psi_i)]  \right\}\,;\\
V_\text{NB} & = & \sum_{i= 1}^{N-3} \sum_{j= i+3}^{N}
V_{ij}(|{\vec{r}}_{i,j}|)\,,
\eea
$\vec{r}_i$ is the position vector of the $i$-th monomer, 
$\vec{r}_{i,j} = \vec{r}_i - \vec{r}_j$, $\vartheta_i$ 
is the $i$-th bond angle, i.e., the 
angle between $\vec{r}_{i+1}$ and $\vec{r}_{i}$, $\psi_i$ the $i$-th dihedral
angle, that is the angle between the vectors $\vec{n}_i = \vec{r}_{i+1,i} \times
\vec{r}_{i+1,i+2}$ and  $\vec{n}_{i+1} = \vec{r}_{i+2,i+1} \times
\vec{r}_{i+2,i+3}$, $k_r = 100$, $a = 1$, $k_\vartheta = 20$, $\vartheta_0 =
105^\circ$, $A_i=0$ and $B_i = 0.2$ if at least two among the residues
$i,i+1,i+2,i+3$ are N, $A_i = B_i = 1.2$ otherwise. As to $V_{ij}$, we have 
$V_{ij} = \frac{8}{3}\left[\left(\frac{a}{r} \right)^{12} + \left(\frac{a}{r}
\right)^{6} \right]$ if $i,j = \text{P},\text{P}$ or $i,j = \text{P},\text{H}$,  
$V_{ij} = 4\left[\left(\frac{a}{r} \right)^{12} - \left(\frac{a}{r}
\right)^{6} \right]$ if $i,j = \text{H},\text{H}$ 
and  $V_{ij} = 4\left(\frac{a}{r}
\right)^{6}$ if either $i$ or $j$ are N
\cite{Thirumalai}. 

Although the identity between dynamical trajectories 
and geodesics of the Eisenhart metric
only holds if the dynamics is the Newtonian one, 
a Langevin dynamics, obtained by adding to the deterministic force $\nabla V$ a 
random force according to the
fluctuation-dissipation theorem and a friction term proportional 
to the velocity,
is a more reasonable model of the dynamics of a polymer in aqueous solution
when the solvent
degrees of freedom are not taken into account explicitly. Since we are
interested not in the details of the time series of $K_R$ along a particular
trajectory but only in its
statistical distribution, we may expect that also a sampling obtained using the
Langevin dynamics gives the same information on the geometry of the landscape. 
To check this assumption 
we let the system evolve with both a
newtonian dynamics (using a symplectic algorithm \cite{MacLachlan} to integrate
the equations of motion) and a
Langevin dynamics (using the same algorithm -- a modified Verlet -- 
and parameters as in Ref.\ \cite{Thirumalai}) obtaining very similar results
in the two cases. In the
following we shall refer only to results obtained with Langevin dynamics.

We considered five different sequences: four of 
22 monomers $S^{22}_{\text{g}} =  \text{P} \text{H}_9 (\text{N}\text{P})_2
\text{N}\text{H}\text{P} \text{H}_3 \text{P} \text{H}$, 
$S^{22}_{\text{b}} = \text{P}\text{H}\text{N}\text{P} \text{H}_3
\text{N}\text{H}\text{N} \text{H}_4 (\text{P}\text{H}_2)_2 \text{P}\text{H}$, 
$S^{22}_{\text{i}} = \text{P}_4 \text{H}_5 \text{N}\text{H} \text{N}_2
\text{H}_6 \text{P}_3$, 
$S^{22}_{\text{h}} = \text{H}_{22}$ 
and also a homopolymeric sequence of 44 monomers
$S^{44}_{\text{h}} = \text{H}_{44}$. 
Sequence $S^{22}_{\text{g}}$ had already 
been identified as a
good folder \cite{Thirumalai} and our simulations confirmed this finding: below
a given temperature it always reached the same $\beta$-sheet-like structure.
Homopolymers $S^{22}_{\text{h}}$ and $S^{44}_{\text{h}}$, on the other hand, 
showed a hydrophobic collapse but no tendency to reach a
particular configuration in the collapsed phase. Sequence $S^{22}_{\text{b}}$
(which has the same overall composition of $S^{22}_{\text{g}}$ rearranged in a
different sequence) behaved as a bad folder and did not reach 
a unique native state, while
$S^{22}_{\text{i}}$ was constructed by us to show a somehow intermediate
behavior between good and bad folders: it always formed the same structure
involving the middle of the sequence,
while the beginning and the end of the chain fluctuated
also at low temperature.
As to standard thermodynamic observables, all the sequences showed very similar
behaviors: in particular, both the specific heat $c_V$ of the
homopolymer $S^{22}_{\text{h}}$ and of the good folder $S^{22}_{\text{g}}$ 
exhibit a peak at the transition (data not shown), and on the sole 
basis of this quantity it would be
hard to discriminate between a simple hydrophobic collapse and a folding. 

\begin{figure}
\begin{center}
\includegraphics[width=12cm,clip=true]{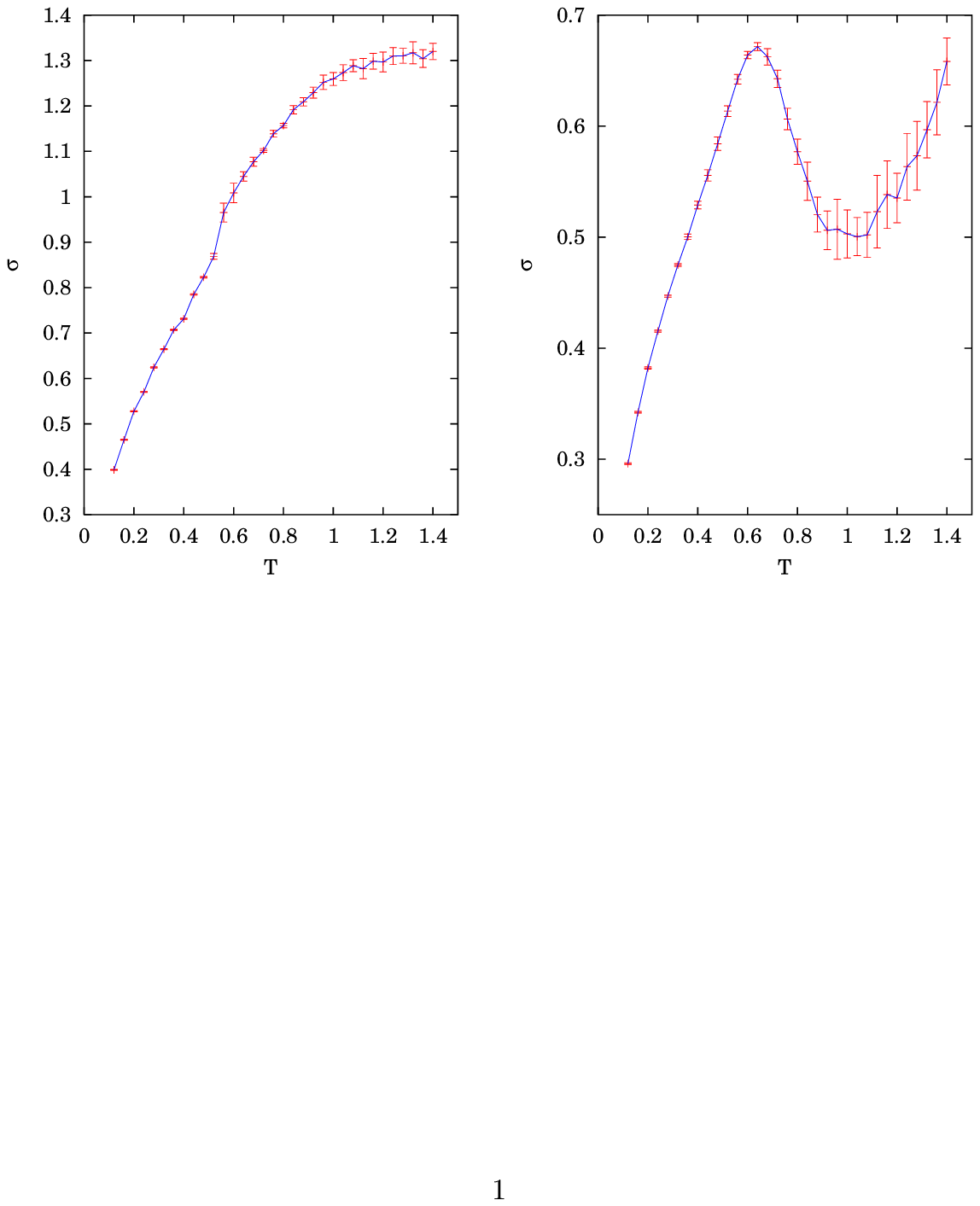}
\end{center}
\caption{Relative curvature fluctuation $\sigma$ vs.\ temperature $T$ for the
homopolymer $S^{22}_{\text{h}}$ (left) and for the good folder
$S^{22}_{\text{g}}$ (right). The solid curves are a guide to the eye.}
\label{figure_sigma}
\end{figure}

On the other hand, a dramatic difference between the homopolymer and the good
folder shows up if we consider the geometric properties of the landscape, and in
particular the fluctuations of the Ricci curvature $K_R$ (\ref{ricci}). 
We defined a relative adimensional curvature fluctuation
$\sigma$ as
\beq
\sigma = \frac{\sqrt{\frac{1}{N}\left( \langle K_R^2\rangle_t - \langle
K_R\rangle_t^2\right)}}{\frac{1}{N}\langle K_R\rangle_t}
\eeq
where $\langle \cdot \rangle_t$ stands for a time average: 
in Fig.\ \ref{figure_sigma} we plot $\sigma$ as a function of the 
temperature $T$
for the homopolymer $S^{22}_{\text{h}}$ and for the good folder 
$S^{22}_{\text{g}}$. A peak shows up in the case of the good folder, close
to the folding temperature $T_f$ (which we estimated as $T_f = 0.6 \pm 0.05$), 
below which the system is mostly in the native state, while
no particular mark of the hydrophobic collapse can be seen in the case of the
homopolymer. As to the other sequences, for the longer homopolymer
$S^{44}_{\text{h}}$ $\sigma(T)$ is even smoother than for $S^{22}_{\text{h}}$, 
at variance with the specific heat which develops a sharper peak consistently
with the presence of a thermodynamic $\theta$-transition as $N \to \infty$ 
(data not shown); for the bad folder $S^{22}_{\text{b}}$, $\sigma(T)$ is not as
smooth as for the homopolymers, but only a very weak signal is found at a lower
temperature than that of the peak in $c_V$, i.e., at the temperature 
where the system starts to behave as
a glass; for the ``intermediate'' sequence $S^{22}_{\text{i}}$ a peak is present
at the ``quasi-folding'' temperature, although considerably broader than in the
case of $S^{22}_{\text{g}}$ (data not shown). 

The behavior of $\sigma(T)$ can thus be used to mark the folding transition and
to identify good folders within the model considered here. It must be stressed
that no knowledge of the native state is necessary to define $\sigma$, and that
it can be computed with the same computational effort needed to obtain the
specific heat and other thermodynamic observables. 

Which is the origin of this behavior of $\sigma(T)$? While we do not have
a complete answer yet, we argue it is a consequence of the effective two-state
dynamics of this system close to the folding transition: an
average curvature in the folded state
considerably larger than in the denatured state (due to the more pronounced
effective potential well in the native state) would naturally imply a
sudden increase of the fluctuations of the curvature as the system approaches
the folding transition. Moreover, higher fluctuations imply a higher degree of 
instability of the dynamics, as
is expected close to $T_f$ where the polymer has essentially the same
probability of being folded or swollen. This result also 
opens a connection between the folding transition and
symmetry-breaking phase transitions: the behavior of $\sigma(T)$  observed here
for the good folder $S^{22}_{\text{g}}$ is remarkably close to that exhibited
by finite systems undergoing a symmetry-breaking phase transition in the
thermodynamic limit \cite{TdF_geometry}. This suggests
that the folding of a proteinlike heteropolymer does share some features of
``true'' symmetry-breaking phase transitions, at least those that show up
already in finite systems, although no singularity in the thermodynamic limit
occurs, because proteins are intrinsically finite objects
\cite{Dill,CecconiBurioni}.  

To summarize, we have shown that the geometry of the energy landscape, and in
particular the fluctuations  $\sigma$ of its curvature,  
can be used to mark the folding transition and to identify 
polymers having a protein-like behavior, in the context of a minimalistic model. 
If tested successfully on
other, maybe more refined models of proteins, $\sigma$ 
might prove a useful tool in
the search of protein-like sequences.
The geometric nature of $\sigma$ may
provide an insight into the nature of the folding transition itself and suggests
a connection with symmetry-breaking phase transitions.

We thank Lorenzo Bongini and Aldo Rampioni for useful discussions and
suggestions. This work is part of the EC FP6 {\em EMBIO} project 
(EC contract n.\ 012835).


\begin{thebibliography}{99}
\bibitem{Anfinsen} C.\ Anfinsen, Science {\bf 181}, 223 (1973).
\bibitem{proteinbook} A.\ Finkelstein and O.\ B.\ Ptitsyn, {\em Protein physics}
(Academic Press, London, 2002).
\bibitem{Wales} D.\ J.\ Wales, 
{\em Energy landscapes} (Cambridge University Press, Cambridge, 2004).
\bibitem{naturecomplex} P.\ G.\ Debenedetti and F.\ H.\ Stillinger, Nature {\bf
410}, 259 (2001).
\bibitem{funnel} J.\ N.\ Onuchic, P.\ G.\ Wolynes, Z.\ Luthey-Schulten, 
and N.\ D.\ Socci, Proc.\ Natl.\ Acad.\ Sci.\ USA {\bf 92}, 3626 (1995).  
\bibitem{numrec} W.\ H.\ Press, S.\ A.\ Teukolsky, W.\ T.\ Vetterling, 
and B.\ P.\ Flannery, {\em Numerical recipes} (Cambridge University Press,
Cambridge, 1992).
\bibitem{Grigera} T.\ S.\ Grigera, A.\ Cavagna, I.\ Giardina, and G.\ Parisi,
Phys.\ Rev.\ Lett.\ {\bf 88}, 055502 (2002).
\bibitem{Clementi} P.\ Das, S.\ Matysiak, and C.\ Clementi, 
Proc.\ Natl.\ Acad.\ Sci.\ USA {\bf 102}, 10141 (2005).
\bibitem{Miller_Wales} M.\ A.\ Miller and D.\ J.\ Wales, J.\ Chem.\ Phys.\ {\bf
111}, 6610 (1999).
\bibitem{Bongini1} L.\ Bongini, R.\ Livi, A.\ Politi, and A.\ Torcini, 
Phys.\ Rev.\ E {\bf 68}, 061111 (2003).  
\bibitem{Bongini2} L.\ Bongini, R.\ Livi, A.\ Politi, and A.\ Torcini, 
Phys.\ Rev.\ E {\bf 72}, 051929 (2005).  
\bibitem{geometry} See e.g.\ M.\ P.\ do Carmo, {\em Riemannian geometry}
(Birkhauser, Boston, 1992), or M.\ Nakahara, {\em Geometry, Topology and
Physics} (IOP Publishing, London, 2003).
\bibitem{Arnold} V.\ I.\ Arnol'd, {\em Mathematical methods of classical
mechanics} (Springer, New York, 1989).
\bibitem{prl95pre96} L.\ Casetti, R.\ Livi, and M.\ Pettini,  
Phys.\ Rev.\ Lett.\ {\bf 74}, 375 (1995); L.\ Casetti, C.\ Clementi, and M.\
Pettini, Phys.\ Rev.\ E {\bf 54}, 5969 (1996).  
\bibitem{physrep} L.\ Casetti, M.\ Pettini, and E.\ G.\ D.\ Cohen, 
Phys.\ Rep.\ {\bf 337}, 237 (2000).
\bibitem{Eisenhart} L.\ P.\ Eisenhart, Ann.\ Math.\ {\bf 30}, 591 (1929).
\bibitem{Thirumalai} T.\ Veitshans, D.\ Klimov, and D.\ Thirumalai, 
Fold.\ Des.\ {\bf 2}, 1 (1997).
\bibitem{MacLachlan} P.\ MacLachlan and P.\ Atela, Nonlinearity {\bf 5}, 541
(1992). 
\bibitem{TdF_geometry} L.\ Caiani, L.\ Casetti, C.\ Clementi, and M.\ Pettini,  
Phys.\ Rev.\ Lett.\ {\bf 79}, 4361 (1997); 
L.\ Caiani, L.\ Casetti, C.\ Clementi, G.\ Pettini, 
M.\ Pettini, and R.\ Gatto, Phys.\ Rev.\ E {\bf 57}, 3886 (1998); 
R.\ Franzosi, L.\ Casetti, L.\ Spinelli, and M.\ Pettini,  
Phys.\ Rev.\ E {\bf 60}, R5009 (1999). 
\bibitem{Dill} K.\ A.\ Dill, Biochemistry {\bf 29}, 31 (1990).
\bibitem{CecconiBurioni} R.\ Burioni, D.\ Cassi, F.\ Cecconi, and A.\ Vulpiani,
Proteins {\bf 55}, 529 (2004).

\end{thebibliography}
\end{document}